\documentclass[%
twocolumn,
 amsmath,amssymb,
 aps, 
pra,
]{revtex4-2}
\usepackage{graphicx}% Include figure files
\usepackage{dcolumn}% Align table columns on decimal point
\usepackage{bm}% bold math
\usepackage{ulem}

\usepackage{enumitem}
\usepackage{csquotes}
\usepackage{appendix}
\usepackage{multirow}
\usepackage[colorlinks=true, allcolors=blue]{hyperref}

\begin{document}

\title{\textbf{Structural perturbation theory for bound states in the continuum via bifurcation of zeros and extrema of the dispersion relation}}

\author{Lijun Yuan}
\email{ljyuan@ctbu.edu.cn}
\affiliation{
 School of Mathematics and Statistics, Chongqing Technology and Business University, Chongqing 400067, China
}%

\author{Ya Yan Lu}
\affiliation{%
 Department of Mathematics, City University of Hong Kong, Kowloon, Hong Kong, China
}%

\begin{abstract}

In a lossless periodic structure, a bound state in the continuum (BIC) corresponds to a real zero and a local maximum of the imaginary part of a complex dispersion relation $k = k(\beta)$, where $\beta$ is the Bloch wave number. A perturbation of the structure deforms the dispersion curve and may destroy, move or split the BIC, as demonstrated in existing studies involving lossless symmetry-preserving perturbations, symmetry-breaking perturbations and dissipative perturbations. We present a comprehensive perturbation theory, emphasizing the evolution of
the real zeros and extreme points of $\mathrm{Im}[k(\beta)]$ under various structural perturbations. In particular, our theory reveals the existence of  lasing threshold modes (LTMs), which are also zeros of $\mathrm{Im}[k(\beta)]$ when the perturbation involves gain with or without balanced loss. Using local Taylor expansions and Puiseux series, we determine the number, locations, and leading-order scaling of real zeros and extreme points for various types of BICs  under different types of structural perturbations. The theory recovers known results and predicts new behavior for super-BICs under  $\mathcal{PT}$-symmetric perturbations. Specifically, a  propagating super-BIC corresponding to a fourth-order zero of $\mathrm{Im}[k(\beta)]$ splits into two real zeros representing either two  BICs or two LTMs, and  a symmetric standing wave splits into two BIC-LTM pairs. Our theory provides a general framework for studying BICs and nearby resonant modes in both lossless and non-Hermitian periodic structures.

\end{abstract}

\maketitle

\section{Introduction}

Bound states in the continuum (BICs)  are spatially localized modes with frequencies inside the radiation continuum~\cite{mari08,hsu16,kosh19}. In recent years,
BICs have attracted sustained interest  in the photonics community, due to their numerous applications including low-threshold lasing~\cite{kodi17,kosh20,ren22,do25}, enhanced sensing~\cite{yesi19,luo24}, harmonic generation, and wave control in non-Hermitian systems~\cite{zele24,vale25,azzam,kang23}.
In practice,  ideal BICs are inevitably subject to perturbations, which may originate from variations in solution parameters or from structural modifications of the system.
In periodic structures, the variation of Bloch wave number $\beta$ gives rise to   a family of resonant modes whose quality factor, $Q$ factor, scales as $(\beta-\beta_*)^{-2n}$, where  $\beta_*$  is the Bloch wave number of  the BIC. Typical BICs have $n=1$, and there are also many super-BICs with $n \geq 2$~\cite{yuan17pra,yuan18,jin19,yuan20spbic,kang21,bul23,le24,zhang25}.

The effect of structural perturbations is more complicated and depends on the order $n$ of the BIC and the type of perturbation. Existing studies revealed that, for lossless symmetry-preserving (lossless-SP) perturbations that maintain the symmetry in the periodic direction, certain generic BICs are robust \cite{yuan17,yuan21}. For balanced loss and gain (i.e., $\mathcal{PT}$-symmetric) perturbations,  a generic BIC typically splits into a BIC and a lasing threshold mode (LTM)~\cite{kar18,han20,song25}. For lossless symmetry-breaking (lossless-SB), dissipative, and gain perturbations, BICs are expected to disappear \cite{yuan20spbic,hu18,kosh18,rozman24,zhou24,sem26,gan20}. However, symmetry-preserving dissipative perturbations can give rise to a symmetry-protected complex BIC (cBIC) with a complex frequency and an infinite radiative $Q$ factor \cite{hu20,song23}.  
Gain perturbations can induce LTMs, whose complex conjugates correspond to coherent perfect absorption (CPA) states ~\cite{chong10,midya18}. To the best of our knowledge, there is no quantitative theory to predict the locations of the resonant modes with the highest $Q$ factor and the locations of the LTMs. For super-BICs, bifurcation theories have been established for lossless-SP perturbations~\cite{zhang24a,zhang24b,zhang25}, showing that they may split into multiple generic BICs or disappear. The evolution of super-BICs under other types of perturbations  has not been investigated.

In a lossless periodic structure with a BIC, resonant modes with $\beta$ near $\beta_*$ satisfy a complex dispersion relation $k = k(\beta)$, where $k = \omega / c$ is the free-space wave number. 
In this work, we analyze the evolution of the  dispersion relation under structural perturbations, emphasizing the imaginary part $\mathrm{Im}[k(\beta)]$. We are interested in tracking the real zeros and extreme points of $\mathrm{Im}[k(\beta)]$. In lossless periodic structures,  a real zero corresponds to a BIC; in non-Hermitian structures, it can be  either a BIC or an LTM. Using local Taylor expansions  and Puiseux series, we determine the number, locations, and leading-order scaling of the real zeros and extreme points of $\mathrm{Im}(k)$. Our theory is applicable to any type of BIC and arbitrary structural perturbations. For super-BICs under $\mathcal{PT}$-symmetric perturbations, we find a number of new wave phenomena. In particular, we show that a propagating super-BIC with $n=2$ yields two zeros that are both BICs or both LTMs (rather than a single BIC-LTM pair), and a symmetric standing wave (SSW) splits into two BIC-LTM pairs.
%We also find that the scaling exponent of the real zeros and extreme point with respect to the perturbation amplitude are different for different types of BICs.

The remainder of this paper is organized as follows. 
In Section~\ref{sec:formulation}, we introduce the  mathematical formulation and basic concepts.
In Section~\ref{sec:theory}, we present a procedure based on Puiseux series for analyzing real zeros and extreme points of $\mathrm{Im}(k)$.
In Section~\ref{sec:result}, we apply the method to five types of BICs and validate the theory numerically. The paper is concluded with a brief discussion in Section~\ref{sec:conclusion}.

\section{Problem Formulation}
\label{sec:formulation}

We consider a two-dimensional lossless structure that is invariant in  $x$, periodic in  $y$ with period $L$, and sandwiched between two homogeneous media in the $z$ direction. For simplicity, we assume the homogeneous media are vacuum.  In addition, we assume that the structure has reflection symmetries in both $y$ and $z$ directions. The dielectric function $\epsilon_*$ is real and satisfies 
$$\epsilon_*(\mathbf{r}) = \epsilon_*(y+L,z) = \epsilon_*(-y,z) = \epsilon_*(y,-z)$$
for all $\mathbf{r} = (y,z)$, and $\epsilon_*(\mathbf{r}) = 1$ for $|z| > d$, where $2d$ is the thickness of the periodic layer.

We further assume that the structure supports a BIC $u_*(\mathbf{r}) = \phi_*(\mathbf{r}) e^{{\sf i} \beta_* y}$ with free-space wave number $k_*=\omega_*/c$ and Bloch wave number $\beta_*$, where $\phi_*$ is periodic in $y$ with period $L$, and  $\omega_*$ is the angular frequency satisfying
\begin{equation}
\label{eq:one_channel}
|\beta_*| < k_*  <  \frac{2\pi}{L} - |\beta_*|.
\end{equation}
In that case, only the zeroth diffraction order is open. We normalize the BIC wave field by
\(
\int_{\Omega}\epsilon_*|\phi_*|^2\,d\mathbf r=1,
\)
where $\Omega = \{(y,z): |y| < L/2, |z| < \infty \}$.

In the vicinity of the BIC, there exists a family of resonant modes characterized by a complex dispersion relation $k = k(\beta)$, where $k$ is the free-space wave number and $\beta$ is the Bloch wave number. The imaginary part of the dispersion relation, $\mathrm{Im}[k(\beta)]$, vanishes at $\beta=\beta_*$ and is tangential to the real axis at that point. More precisely, for $\beta$ near $\beta_*$,
\begin{equation}
\label{eq:disp_BIC}
\mathrm{Im}[k(\beta)] \propto - (\beta-\beta_*)^{2n},
\end{equation}
where $n$ is a positive integer characterizing the order of the BIC. A typical BIC has $n=1$. If $n\geq 2$, the BIC is referred to as a super-BIC.

We consider a structural perturbation by assuming that the dielectric function takes the form
\begin{equation}
\label{eq:perturbation}
\epsilon(\mathbf{r};\delta)=\epsilon_*(\mathbf{r})+\delta F(\mathbf{r}),
\end{equation}
where  $F(\mathbf{r})$ is the perturbation profile and $\delta$ is the perturbation amplitude. In the above, $F(\mathbf{r})$ is  an $O(1)$ function, even in $z$, periodic in $y$ with period $L$, and vanishes for $|z|>d$. Consequently, the perturbation preserves the periodicity of the structure and does not change the surrounding homogeneous media.
We consider five types of structural perturbations:
lossless symmetry-preserving (Lossless-SP) perturbation,
lossless symmetry-breaking (Lossless-SB) perturbation,
gain perturbation,
dissipative perturbation,
and balanced loss-and-gain  ($\mathcal{PT}$-symmetric) perturbation. The corresponding perturbation profiles are specified below:
\begin{itemize}[leftmargin=2cm]
    \item[Lossless-SP]: $F$ is real and satisfies $F(\mathbf{r}) = F(-y,z)$;
    \item[Lossless-SB]: $F$ is real and satisfies $F(\mathbf{r}) \not\equiv F(-y,z)$;
    \item[Dissipative]: $F$ is purely imaginary with $\mathrm{Im}(F)\ge0$;
    \item[Gain]: $F$ is purely imaginary with $\mathrm{Im}(F)\le0$;
    \item[$\mathcal{PT}$]: $F$ is purely imaginary and satisfies $F(\mathbf{r}) = - F(-y,z)$.
\end{itemize}
These perturbation profiles represent the main physically distinct and highly relevant scenarios in practical applications.
%, including dissipation, optical gain, $\mathcal{PT}$-symmetry, and symmetric lossless perturbations.

In the perturbed structure, resonant modes depend on both the Bloch wave number $\beta$ and the perturbation amplitude $\delta$. Accordingly, the wave field $\phi(\mathbf{r})=\phi(\mathbf{r};\beta,\delta)$ and the associated complex wave number $k=k(\beta,\delta)$ are functions of these two parameters. The resonant mode satisfies the governing equation
\begin{equation}
\label{eq:resonantmode}
\mathcal{L}(\beta,\delta)\,\phi(\mathbf{r})=0,
\end{equation}
where 
\begin{equation}
\label{eq:operator_L}
\mathcal{L}(\beta,\delta)
=
\partial^2_y + \partial^2_z 
+2{\sf i}\beta\,\partial_y
+k^2(\beta,\delta)\epsilon(\mathbf{r}; \delta)
-\beta^2.
\end{equation}
The BIC of the unperturbed structure corresponds to the resonant mode at $(\beta,\delta)=(\beta_*,0)$, with field $\phi_*(\mathbf{r})=\phi(\mathbf{r};\beta_*,0)$ and wave number $k_*=k(\beta_*,0)$. It satisfies
$  \mathcal{L}_*\,\phi_*(\mathbf{r})=0, $  where $ \mathcal{L}_*=\mathcal{L}(\beta_*,0). $ 
At $\delta=0$,  $\beta_*$ is a real zero and an extreme point of $\mathrm{Im}[k(\beta,0)]$.

Reciprocity gives rise to
\begin{equation}
\label{eq:reciprocity}
k(\beta, \delta) = k(-\beta, \delta)
\end{equation}
for  $\beta \in (-\pi/L, \pi/L]$ and real $\delta$. For $\mathcal{PT}$-symmetric perturbations, it is easy to check that 
$$\epsilon(-y,z; \delta) = \epsilon(y, z; -\delta).$$ Thus, the dispersion relation satisfies $k(\beta, \delta) = k(-\beta, -\delta)$. In addition,  Eq.~(\ref{eq:reciprocity}) and the above relation imply 
\begin{equation}
    \label{eq:PT_reciprocity}
k(\beta, \delta) = k(\beta, -\delta)
\end{equation}
for  $\beta \in (-\pi/L, \pi/L]$ and real $\delta$.

\section{Perturbation method}
\label{sec:theory}
In this section, we use Puiseux series to study  the real zeros and extreme points of $\mathrm{Im}(k)$ for $\delta > 0$. We start with the  Taylor expansion of $\mathrm{Im}(k)$ at $(\beta_*, 0)$, i.e., 
\begin{equation}
\label{eq:TaylorExpImK}
\mathrm{Im}[k(\beta, \delta)] = \sum_{j=0}^{\infty}\sum_{m=0}^{\infty}
a_{jm} (\beta - \beta_*)^{j} \delta^{m},
\end{equation}
where 
$$a_{jm}=\frac{1}{j!m!}\,\mathrm{Im}\!\left(k_{jm}\right),
\quad
k_{jm}=
\frac{\partial^{j+m}k}{\partial\beta^{j}\partial\delta^{m}}(\beta_*,0).$$ 
Clearly,  $k_{00} = k_*$ is real and $a_{00} = 0$. For other pairs $(j,m)$, $k_{jm}$ and $a_{jm}$ can be obtained iteratively by differentiating the governing equation (\ref{eq:resonantmode}) with respect to $\beta$ and $\delta$, and enforcing solvability conditions at each order. In practice, only a few lowest-order terms are required.  

It is easy to see that  $a_{j,0} = 0$ for $j < 2n$, where $n$ is the order of the BIC. For a BIC with $\beta_* = 0$, according to  Eq.~(\ref{eq:reciprocity}), we have $a_{jm} = k_{jm} = 0$ for odd $j$ and all $m \geq 0$. For any BIC, if the perturbation is $\mathcal{PT}$-symmetric, then Eq.~(\ref{eq:PT_reciprocity}) leads to $a_{jm} = k_{jm} = 0$ for all $j \geq 0$ and odd $m $. The  key coefficients $a_{jm}$ are listed in Appendix~A for different types of BICs and different perturbation profiles $F$.

For a given small $\delta>0$, we construct the zeros of $\mathrm{Im}(k)$ near $\beta_*$ using Puiseux series as
\begin{equation}
\label{eq:exp_beta_zero}
\beta^{\mathrm{Zero}}(\delta)
=
\beta_*
+\beta^{\mathrm{Zero}}_1\,\delta^{\alpha}
+\beta^{\mathrm{Zero}}_2\,\delta^{2\alpha}
+\cdots,
\end{equation}
where $\alpha > 0$ is the exponent to be determined.  If real coefficients $\beta^{\mathrm{Zero}}_j$ for $j \geq 1$ exist such that $\mathrm{Im}\{k[\beta^{\mathrm{Zero}}(\delta),\delta]\}=0$ for $\delta > 0$, then expansion (\ref{eq:exp_beta_zero}) yields a real zero.   Substituting expansion 
\eqref{eq:exp_beta_zero} into  (\ref{eq:TaylorExpImK})  and matching the
lowest-order terms in $\delta$, we can determine $\alpha$ and obtain an algebraic equation for $\beta^{\mathrm{Zero}}_1$. The number of real roots of this algebraic equation gives the number of real zeros.  After $\beta_1^{\sf Zero}$ is calculated, higher-order coefficients $\beta_j^{\sf Zero}$ for $j \geq 2$ can be obtained iteratively by balancing the terms of $\delta^{j \alpha}$. In Appendix A, we show how to compute $\alpha$ and $\beta_1^{\sf Zero}$ for a propagating BIC with $n=1$.

Similarly, the extreme points of $\mathrm{Im}(k)$  are constructed as
\begin{equation}
\label{eq:exp_beta_extreme}
\beta^{\mathrm{Ext}}(\delta)
=
\beta_*
+\beta^{\mathrm{Ext}}_1\,\delta^{\alpha}
+\beta^{\mathrm{Ext}}_2\,\delta^{2\alpha}
+\cdots.
\end{equation}
The real coefficients $\beta^{\mathrm{Ext}}_j$ for $j \geq 1$ are obtained by requiring $\mbox{Im}\{ \partial_{\beta} k[ \beta^{\sf Ext}(\delta), \delta] \} = 0.$
We denote the resulting local maximum and minimum points by $\beta^{\mathrm{Max}}$ and $\beta^{\mathrm{Min}}$, respectively. The corresponding free-space wave numbers are denoted by $k^{\sf Zero}, k^{\sf Ext}, k^{\sf Max}$ and $k^{\sf Min}$.

\section{Results}
\label{sec:result}

In this section, we present the theoretical results  obtained using the perturbation method in Section~\ref{sec:theory}, and validate the results  by numerical examples. 
We start with generic propagating BICs of order $n=1$, then move on to super-BICs with $n=2$, including both propagating and  standing-wave cases, and finally consider antisymmetric standing waves (ASWs) with $n=3$. Unless otherwise stated, all zeros mentioned below refer to real zeros in a neighborhood of the unperturbed BIC.

\subsection{Generic propagating BICs}

For a  propagating BIC with $n=1$,  $\beta_* \neq 0$ is a double zero and a local maximum point of $\mathrm{Im}[k(\beta, 0)]$. We have $a_{10}=0$ and $a_{20} < 0$. For $\delta>0$, the number and locations of the zeros and extreme points of $\mathrm{Im}(k)$ for different types of perturbations are summarized in Table~\ref{tab:PBIC}. 
We observe that for all perturbation types, a nearby local maximum 
persists continuously for sufficiently small positive $\delta$.
In contrast, the existence of a zero depends on the perturbation profile $F$.
It may persist, disappear, or split into two distinct zeros.

For lossless-SP perturbations, a double zero remains.
The zero still coincides with the local maximum point, and it  corresponds to a BIC.
This indicates that the BIC is robust under such perturbations,
consistent with the theory developed in~\cite{yuan17}.
For lossless-SB perturbations, 
the  local maximum satisfies 
$
\mathrm{Im}(k^{\sf Max}) \sim
\eta_1 \delta^2,
$
where $\eta_1  < 0$ as shown in Appendix A. Thus, $\mathrm{Im}(k^{\sf Max}) < 0$ for small $\delta$.
As a result, no zero exists in this case.

For dissipative and gain perturbations, the local maximum point satisfies $\beta^{\sf Max} - \beta_* \sim - ({0.5 a_{11}}/{ a_{20}}) \delta$ 
and $\mbox{Im}(k^{\sf Max}) \sim a_{01} \delta,$ where $a_{01} = - k_* \mbox{Im} \left[ \int_{\Omega} F | \phi_*|^2 d \mathbf{r} \right] / 2$. 
If the perturbation is dissipative and the perturbation region has a nonzero overlap with the BIC field, then  $a_{01} < 0$ and $\mbox{Im}(k^{\sf Max}) < 0$ for small positive $\delta$. Hence,  the local maximum lies below the real axis, and there is no zero. Notice that $\mathrm{Im}(k^{\sf Max})$ has a different order in $\delta$ for dissipative and lossless-SB perturbations. Therefore, although both perturbations destroy the BIC, the $Q$ factor of the resulting resonant modes scales differently with $\delta$. More precisely, the $Q$ factor is $O(1/\delta)$ and $O(1/\delta^2)$ for dissipative and lossless-SB perturbations, respectively.

If the perturbation is gain, then $a_{01} > 0$ and $\mbox{Im}(k^{\sf Max}) > 0$. Thus,  $\mathrm{Im}(k)$ is shifted upward and crosses the real axis at two distinct Bloch wave numbers. This gives rise to two simple zeros. In non-Hermitian systems, a  zero of $\mathrm{Im}(k)$ for $\delta > 0$ corresponds either to a BIC or to an LTM. In Appendix B, we present a procedure to distinguish these two possibilities by examining the asymptotic behavior of the associated wave field.
For this case, we can show that the two zeros correspond to two LTMs.
We denote the Bloch wave number and free-space wave number of an LTM by
$\beta^{\sf LTM}$ and $k^{\sf LTM}$. A CPA state can be realized
in a structure with dielectric function
$\bar{\epsilon} = \epsilon_* + \delta \overline{F}$.
It occurs for two counter-propagating incident waves characterized by
$(k^{\sf LTM}, -\beta^{\sf LTM})$.

For $\mathcal{PT}$-symmetric perturbations, the local maximum satisfies
$
\mathrm{Im}(k^{\sf Max}) \sim a_{02}\, \delta^2 > 0.
$
While it is clear that $\mathrm{Im}(k^{\sf Max})$ should be positive and negative for gain and dissipative perturbations, respectively, it is not obvious why $\mathrm{Im}(k^{\sf Max})$ is positive for $\mathcal{PT}$-symmetric perturbations. Nevertheless, we can rigorously prove that $a_{02} > 0$ for generic $\mathcal{PT}$-symmetric perturbations. 
In this case, there are also two zeros.
However, unlike in the case of gain perturbations,
only one zero corresponds to a BIC, while the other zero represents an LTM. 
This is another nontrivial result which we can justify rigorously.

\begin{table}[htbp]
    \centering
\caption{Local maximum point and zeros of $\mathrm{Im}(k)$ for $\delta>0$ near generic propagating BICs. Constant $\eta_1$ is defined in Appendix A. }
\label{tab:PBIC}
    \begin{tabular}{|c|c|c|c|c|c|c|}\hline
        \rule{0pt}{3ex}
         \multirow{2}{*}{$F$}  & \multicolumn{3}{c|}{Maximum point} & \multicolumn{3}{c|}{Zeros} \\ 
         \cline{2-7}
         \rule{0pt}{3ex}
           & $\alpha$  & $\beta^{\sf Max} - \beta_* \sim $ & $\mathrm{Im}(k^{\sf Max}) \sim $ & No. &  $\alpha$ &  $\beta^{\sf Zero} - \beta_* \sim $ \\[2pt] \hline
           \rule{0pt}{3ex}
Lossless-SP & $1$ & $   - \frac{a_{11}}{2a_{20}} \delta   $ & $0$ &   $1$ & $1$ & $   \beta^{\sf Zero} = \beta^{\sf Max} $ \\[3pt] \hline
          \rule{0pt}{3ex}
Lossless-SB  & $1$& $   - \frac{a_{11}}{2a_{20}} \delta   $ & $ \eta_1 \delta^2 $ &  $0$ & $- $ & $-$ \\[3pt] \hline
          \rule{0pt}{3ex}
 Dissipative  & $1$ & $   - \frac{a_{11}}{2 a_{20}} \delta $ &  $ a_{01} \delta $ & $0$& $-$  &  $-$ \\[3pt] \hline
 \rule{0pt}{3.5ex}
 Gain  & $1$ & $   - \frac{a_{11}}{2 a_{20}} \delta   $ & $ a_{01} \delta $ & 2 & $\dfrac{1}{2}$  & $  \pm \sqrt{-\frac{a_{01}}{a_{20}}}  \delta^{\frac{1}{2}}$ \\[4pt] \hline
 \rule{0pt}{3ex}
 $\mathcal{PT}$  & 2 & $   - \frac{a_{12}}{2 a_{20}} \delta^2 $ &  $ a_{02} \delta^2 $ &  $2$ & $1$ & $  \pm \sqrt{  -\frac{a_{02}}{a_{20}} } \delta $ \\[3pt] \hline
 \end{tabular} 
\end{table}

\begin{figure}
    \centering
    \includegraphics[width=0.9\linewidth]{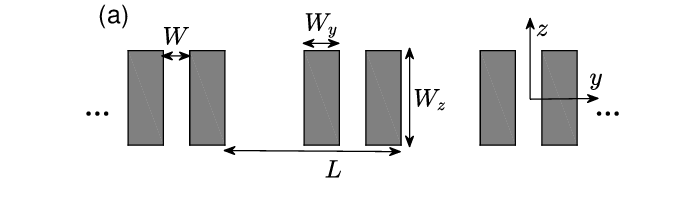} \\
    \includegraphics[width=0.45\linewidth]{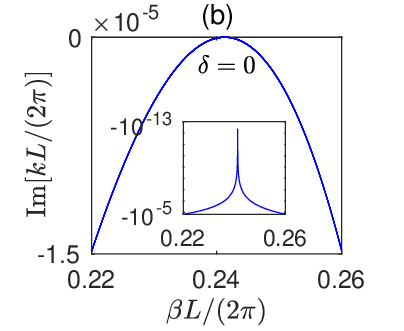}
    \includegraphics[width=0.45\linewidth]{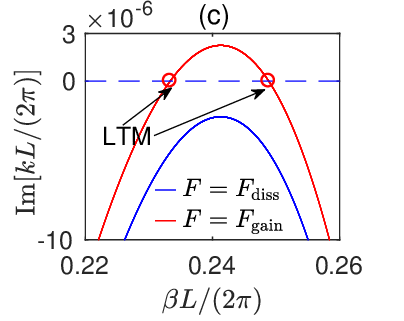} \\
    \includegraphics[width=0.45\linewidth]{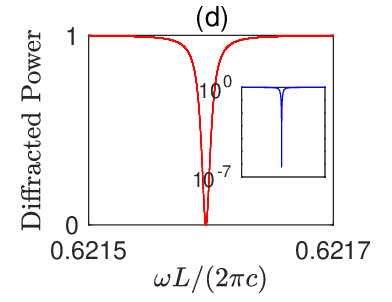}
     \includegraphics[width=0.45\linewidth]{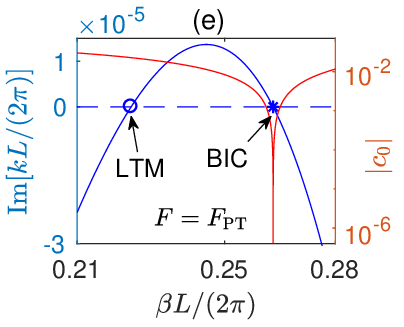}
    \caption{
Evolution of the dispersion relation near the propagating BIC of order $n=1$. (a): The periodic structure.
(b):  $\mathrm{Im}[k(\beta, 0)]$. The inset shows the logarithmic scale. 
(c): $\mathrm{Im}(k)$ for dissipative (blue) and gain (red) perturbations at $\delta=0.0001$. 
(d): Total diffracted power for two incident waves with $\beta = -0.2333\,(2\pi/L)$ illuminating the structure with dielectric function $\overline{\epsilon} = \epsilon_* + \delta \overline{F_{\sf gain}}$. The inset shows the logarithmic scale. 
(e): $\mathcal{PT}$-symmetric perturbation with $\delta=0.1$. Left axis: $\mathrm{Im}(k)$; right axis: radiation coefficient $c_0$ as a function of $\beta$.
}
    \label{fig:genericPBIC}
\end{figure}

To validate the analytical results, we consider  a periodic array of dielectric rectangular cylinders  surrounded by vacuum, as illustrated in Fig.~\ref{fig:genericPBIC}(a). Each period contains two identical cylinders with  dielectric constant $\epsilon_1 = 11.56$. For geometric parameters $W_y = 0.25L$, $W_z = 0.6L$, and $W = 0.2L$, the unperturbed structure supports a generic BIC with $  \beta_* \approx 0.2413 \,(2\pi/L)$, and $  k_* \approx 0.6227 \,(2\pi/L).$ 
We define the perturbation profile $F$ as follows:
\begin{itemize}[leftmargin=2cm]
    \item[Lossless-SP:] 
   $  F_{\sf real}(\mathbf{r}) = \left\{ \begin{matrix} {\sf 1}, & \mathbf{r} \ \ \mbox{in cylinders}, \\
                                            0, & \mbox{otherwise}. \end{matrix} \right.$
    \item[Dissipative:] 
   $  F_{\sf diss}(\mathbf{r}) = \left\{ \begin{matrix} {\sf i}, & \mathbf{r} \ \ \mbox{in cylinders}, \\
                                            0, & \mbox{otherwise}. \end{matrix} \right.$ 
\item[Gain:]  
   $  F_{\sf gain}(\mathbf{r}) = - F_{\sf diss}(\mathbf{r}).$
\item[$\mathcal{PT}$:] 
   $  F_{\sf PT}(\mathbf{r}) = \left\{ \begin{matrix} {\sf i}, & \mathbf{r} \ \ \mbox{in left cylinder}, \\
                                             {\sf -i}, & \mathbf{r} \ \ \mbox{in right cylinder}, \\
                                            0, & \mbox{otherwise}. \end{matrix} \right.$ 
\end{itemize}

Figure~\ref{fig:genericPBIC}(b) shows $\mathrm{Im}[k(\beta,0)]$ as a function of $\beta$, highlighting the double zero at the BIC. 
Fig.~\ref{fig:genericPBIC}(c) shows the effect of small dissipative (i.e., $F = F_{\sf diss}$) and gain (i.e., $F = F_{\sf gain}$) perturbations for $\delta = 0.0001$.  As predicted analytically, for the gain perturbation, $\mathrm{Im}(k)$ intersects the real axis at two distinct Bloch wave numbers, $ \beta^{\sf Zero} \approx 0.2333\,(2\pi/L)$ and $ 0.2488\,(2\pi/L),$  which correspond to LTMs. The complex conjugates of these LTMs represent CPA states.
Fig.~\ref{fig:genericPBIC}(d) illustrates the total diffracted power when two incident plane waves with $\beta \approx -0.2333 \,(2\pi/L)$ illuminate the structure with  dielectric function $\bar{\epsilon} = \epsilon_* + \delta \overline{F_{\sf gain}}$, where the overbar denotes complex conjugation. As expected, complete absorption is observed at the corresponding LTM (CPA) frequency. Fig.~\ref{fig:genericPBIC}(e) (left axis) shows the case of the $\mathcal{PT}$-symmetric perturbation with $\delta = 0.1$. We can see that there are two zeros corresponding to a BIC and an LTM, respectively. We denote the outgoing coefficient of the zeroth diffraction order of a resonant mode by \(c_0\). The distinction between the BIC and the LTM is verified by plotting $c_0$ as a function of $\beta$, shown on the right axis of Fig.~\ref{fig:genericPBIC}(e). The coefficient $c_0$ vanishes at the BIC, but remains finite at the LTM. 
We note that numerical validations for lossless-SP and lossless-SB perturbations are omitted, since these cases have been well studied in the literature.

\subsection{Generic ASWs}

Antisymmetric standing waves with $n=1$ are the simplest BICs in symmetric structures. For an ASW, $\beta_*=0$ and  $\phi_*$ is odd in $y$. 
Near a generic ASW, the numbers of zeros and local maximum points are the same as those of a generic propagating BIC.
However, due to reciprocity, the dispersion relation satisfies $k(\beta,\delta)=k(-\beta,\delta)$ for all real $\beta$ and $\delta$. Therefore,   the local maximum point is always at $\beta=0$. For dissipative and gain perturbations, when the perturbation profile $F$ is even in $y$, the local maximum point at $\beta=0$ corresponds to a cBIC~\cite{hu20}.

\subsection{Second-order propagating super-BICs}
\label{sec:superPBIC}

\begin{table*}[htbp]
    \centering
\caption{Extreme points and zeros of $\mathrm{Im}(k)$ for $\delta>0$ near  propagating super-BICs with $n=2$. Constant $\eta_2$ is defined in Appendix A.}
\label{tab:SuperPBIC}
\begin{tabular}{|c|c|c|c|c|c|c|c|}\hline
    \rule{0pt}{3ex}
         \multirow{2}{*}{$F$}  & \multicolumn{4}{c|}{Extreme points}  & \multicolumn{3}{c|}{Zeros} \\
         \cline{2-8}
        \rule{0pt}{3ex}
           & No. & $\alpha$  & $\beta^{\sf Ext} - \beta_* \sim $ & $\mbox{Im}(k^{\sf Ext}) \sim $ & No.  & $\alpha$ &  $\beta^{\sf Zero} -\beta_* \sim $ \\[2pt] \hline
           \rule{0pt}{4ex}
Lossless-SP & $3$ or $1$ & $\dfrac{1}{2}$ & $  \pm \sqrt{ - \frac{a_{21}}{2a_{40}} } \delta^{\frac{1}{2}},  - \frac{a_{12}}{2a_{21}} \delta  $ & 0,  $a_{02} \delta^2$  &  $2$ or $0$ & $\dfrac{1}{2}$ & $ \beta^{\sf Zero} = \beta^{\sf Max}$ \\[5pt] \hline
\rule{0pt}{4ex}
Lossless-SB & $3$ or $1$ & $\dfrac{1}{2}$ & $  \pm \sqrt{ - \frac{a_{21}}{2a_{40}} } \delta^{\frac{1}{2}},  - \frac{a_{12}}{2a_{21}} \delta $ & $\eta_2 \delta^2$, $a_{02} \delta^2 $  &  $0$ & $-$ & $ - $ \\[5pt] \hline
\rule{0pt}{4ex}
 Dissipative & $1$ & $\dfrac{1}{3}$ & $  - \sqrt[3]{\frac{a_{11}}{4 a_{40}} } \delta^{\frac{1}{3}}  $ & $a_{01} \delta$ & $0$& $-$  &  $-$  \\[5pt] \hline
 \rule{0pt}{4ex}
 Gain  & $1$ & $\dfrac{1}{3}$ & $  - \sqrt[3]{\frac{a_{11}}{4 a_{40}} } \delta^{\frac{1}{3}}   $  &  $a_{01} \delta$ & $2$ & $\dfrac{1}{4}$  & $  \pm \sqrt[4]{-\frac{a_{01}}{a_{40}} }  \delta^{\frac{1}{4}}$ \\[5pt] \hline
 \rule{0pt}{4ex}
 $\mathcal{PT}$  & $1$ & $\dfrac{2}{3}$ & $  - \sqrt[3]{\frac{a_{12}}{4 a_{40}} } \delta^{\frac{2}{3}}  $ & $a_{02} \delta^2$ &  $2$ & $\dfrac{1}{2}$ & $  \pm \sqrt[4]{  -\frac{a_{02}}{a_{40}} } \delta^{\frac{1}{2}} $ \\[5pt] \hline
 \end{tabular}
\end{table*}

In this subsection, we consider propagating super-BICs with $n=2$ and $\beta_* \neq 0$. In that case,  $\beta_*$ is a quadruple zero  of $\mathrm{Im}[k(\beta,0)]$. 
%Thus, $a_{j,0} = 0$ for $j \leq 3$ and $a_{40} < 0$.
%Thus, $a_{j,0}=0$ for $0 \le j \le 3$ and $a_{40}<0$.
The evolution of $\mathrm{Im}(k)$ for $\delta > 0$  under different perturbation types is summarized in Table~\ref{tab:SuperPBIC}.
In contrast to generic propagating BICs,  the  higher-order zero gives rise to multiple extreme points and smaller fractional Puiseux exponents, allowing a single super-BIC to split into multiple generic BICs or disappear, depending on the perturbation profile $F$.

For lossless-SP perturbations, the number of extreme points and zeros depends on the sign of the coefficient $a_{21}$. 
If $a_{21}>0$, then  $\mathrm{Im}(k)$ exhibits three extreme points for small positive $\delta$.
Two of them are local maximum points satisfying
$\beta^{\sf Max}-\beta_* \sim \pm \sqrt{-a_{21}/(2a_{40})}\,\delta^{1/2}$
and $\mathrm{Im}(k^{\sf Max})=0$.
These two maximum points are also zeros and they correspond to BICs. This implies that the super-BIC splits into two generic propagating BICs.
The  local minimum point satisfies
$\beta^{\sf Min}-\beta_* \sim -a_{12}/(2a_{21})\,\delta$ and
$\mathrm{Im}(k^{\sf Min}) \sim a_{02}\delta^2$.
If $a_{21}<0$, there is only one local maximum, and $\mathrm{Im}(k^{\sf Max}) \sim a_{02}\delta^2<0$ for small $\delta$. Thus, no zero exists. 
For lossless-SB perturbations, the results are similar to those for lossless-SP perturbations and likewise depend on the sign of $a_{21}$. 
However, all extreme points satisfy
$\mathrm{Im}(k^{\sf Ext})<0$.
As a result, no zero exists.  Note that $a_{21}$ changes sign when $F$ is replaced by $-F$.

For dissipative and gain perturbations, the results are similar to those of generic propagating BICs, except that the Puiseux exponent $\alpha$ is different.
For a $\mathcal{PT}$-symmetric perturbation, there is a single maximum point satisfying
$\mathrm{Im}(k^{\sf Max}) \sim a_{02}\delta^2>0$ for small $\delta$.
Consequently, $\mathrm{Im}(k)$ exhibits two simple zeros.
However, in contrast to generic propagating BICs, these two zeros correspond either to two BICs or to two LTMs.
Which scenario occurs is determined by the sign of a constant $\chi$, defined in Appendix~B.
If $\chi>0$, both zeros correspond to LTMs, whereas if $\chi<0$, both correspond to BICs. 
Since the unperturbed structure is symmetric with respect to $y$, the existence of a BIC at $(\beta_*,k_*)$ implies a counterpart at $(-\beta_*,k_*)$.
As a result, two additional zeros appear near $-\beta_*$.
If the two zeros near $\beta_*$ correspond to LTMs, those near $-\beta_*$ correspond to BICs, and vice versa.

\begin{figure}
    \centering
    \includegraphics[width=0.45\linewidth]{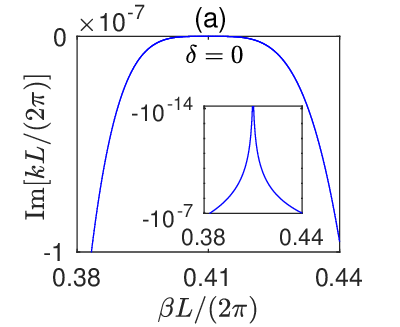}
    \includegraphics[width=0.45\linewidth]{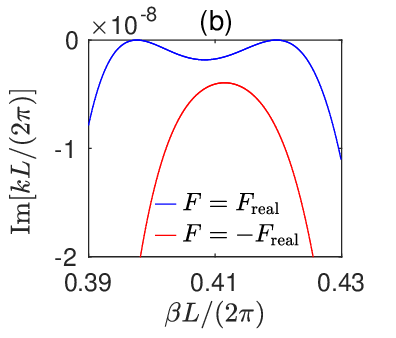} \\
    \includegraphics[width=0.45\linewidth]{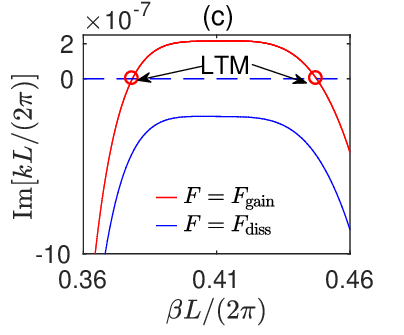} 
     \includegraphics[width=0.45\linewidth]{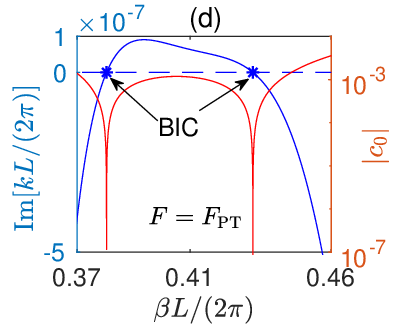} \\
     \includegraphics[width=0.45\linewidth]{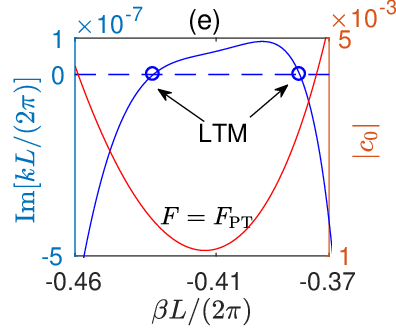}
    \caption{
Evolution of the dispersion relation near the propagating super-BIC. 
(a): $\mathrm{Im}[k(\beta, 0)]$. The inset shows the logarithmic scale. 
(b): $\mathrm{Im}(k)$ for lossless-SP perturbations with $\delta=1$.
(c): $\mathrm{Im}(k)$ for dissipative (blue) and gain (red) perturbations at $\delta=10^{-5}$.  
(d) and (e):  the results for the  $\mathcal{PT}$-symmetric perturbation with $\delta=0.1$ near $\beta_*$ and $-\beta_*$, respectively. Left axis: $\mathrm{Im}(k)$; right axis: radiation coefficient $c_0$ as a function of $\beta$.  }
    \label{fig:SuperPBIC}
\end{figure}

To validate the results listed in Table~\ref{tab:SuperPBIC}, we still consider the periodic structure shown in Fig.~\ref{fig:genericPBIC}(a).
For $W_y=0.33L$, $W_z=1.573355L$, $W=0.1L$, and the same $\epsilon_1$,
the structure supports a propagating super-BIC with $n=2$ at
$\beta_* \approx 0.4096\,(2\pi/L)$ and $k_* \approx 0.5221\,(2\pi /L)$.
Fig.~\ref{fig:SuperPBIC}(a) shows the imaginary part of the dispersion relation near the super-BIC for the unperturbed structure.
%The perturbation profiles $F$ are chosen in the same manner as previous.
Fig.~\ref{fig:SuperPBIC}(b) shows the results for the lossless-SP perturbation with $\delta=1$.
For $F = F_{\sf real}$, we have $a_{21} > 0$,  and there are two double zeros and three extreme points, as shown by the blue curve in Fig.~\ref{fig:SuperPBIC}(b). If we change the sign of $F$, i.e. $F = -F_{\sf real}$, then $a_{21} < 0$, and no zero remains and only a single extreme point survives, as shown by the red curve.

Figure~\ref{fig:SuperPBIC}(c) illustrates the effect of small dissipative and gain perturbations with $\delta=10^{-5}$. For the gain perturbation $F = F_{\sf gain}$, $\mathrm{Im}(k)$ intersects the real axis at $\beta^{\sf Zero} \approx 0.3784\,(2\pi/L)$ and $0.4474\,(2\pi/L)$, corresponding to two LTMs, in agreement with the theoretical predictions. 
Fig.~\ref{fig:SuperPBIC}(d) shows the case of the $\mathcal{PT}$-symmetric perturbation $F = F_{\sf PT}$ and $\delta=0.1$.
Two simple zeros appear, and both correspond to BICs.
Fig.~\ref{fig:SuperPBIC}(e) shows the corresponding results near $-\beta_*$, where the two zeros are LTMs.
We note that the numerical results for a super-BIC under a lossless-SB perturbation are not shown here, since such a perturbation can be regarded as composite: a lossless-SP perturbation first turns the super-BIC into either multiple generic BICs or eliminates it altogether, after which a symmetry-breaking perturbation acts on the resulting generic BICs. Consequently, its qualitative behavior may be understood by comparison with the evolution under lossless-SP perturbations and that of generic BICs under lossless-SB perturbations.

\subsection{Symmetric standing waves}

\begin{table*}[htbp]
    \centering 
\caption{Extreme points and zeros of $\mathrm{Im}(k)$ for $\delta>0$ near  SSWs. Constants $\eta_2 $, $\xi$, $\xi_1$ and $\xi_2$ are defined in Appendix~A.}
\label{tab:SSW}
    \begin{tabular}{|c|c|c|c|c|c|c|c|}\hline
        \rule{0pt}{3ex}
         \multirow{2}{*}{$F$}   & \multicolumn{4}{c|}{Extreme points} &   \multicolumn{3}{c|}{Zeros} \\ 
         \cline{2-8}
         \rule{0pt}{3ex}
           & No. & $\alpha$  & $\beta^{\sf Ext} -\beta_* \sim $ & $\mbox{Im}(k^{\sf Ext}) \sim $  & No.  & $\alpha$ &  $\beta^{\sf Zero} - \beta_* \sim $  \\[2pt] \hline
           \rule{0pt}{4ex}
Lossless-SP & $3$ or $1$ & $\dfrac{1}{2}$ & $  \pm \sqrt{ - \frac{a_{21}}{2a_{40}} } \delta^{\frac{1}{2}}, 0   $ & $0$,  $a_{02} \delta^2$  &   $2$ or $0$ & $\dfrac{1}{2}$ & $  \beta^{\sf Zero} = \beta^{\sf Max} $  \\[5pt] \hline
\rule{0pt}{4ex}
Lossless-SB & $3$ or $1$ & $\dfrac{1}{2}$ & $  \pm \sqrt{ - \frac{a_{21}}{2a_{40}} } \delta^{\frac{1}{2}}, 0 $ & $\eta_2 \delta^2$, $a_{02} \delta^2$ &   $0$ & $-$ & $  - $  \\[5pt] \hline
\rule{0pt}{4ex}
Dissipative  & $3$ or $1$ & $\dfrac{1}{2}$ & $  \pm \sqrt{-\frac{a_{21}}{2 a_{40}} } \delta^{\frac{1}{2}},  0  $ &  $ a_{01} \delta $ & $0$& $-$  &  $-$  \\[5pt] \hline
\rule{0pt}{4ex}
 Gain  & $3$ or $1$ & $\dfrac{1}{2}$ & $  \pm \sqrt{-\frac{a_{21}}{2 a_{40}} } \delta^{\frac{1}{2}}, 0  $ & $ a_{01} \delta $ &  2 & $\dfrac{1}{4}$  & $   \pm \sqrt[4]{-\frac{a_{01}}{a_{40}} }  \delta^{\frac{1}{4}}$ \\[5pt] \hline
 \rule{0pt}{4ex}
 $\mathcal{PT}$  & $3$ or $1$ & $1$ & $   \pm \sqrt{ - \frac{a_{22}}{2 a_{40}} } \delta,  0 $ & $ \xi \delta^4$, $ a_{04} \delta^4 $ & $4$ or $0$ & $1$ & $  \pm \sqrt{  \xi_1 \pm \sqrt{\xi_1^2 - \xi_2}} \delta $ \\[5pt] \hline
 \end{tabular} 
\end{table*}

We now turn to SSWs, which are super-BICs of order $n=2$.
For an SSW, $\beta_*=0$ is a quadruple zero of $\mathrm{Im}[k(\beta, 0)]$,  and  $\phi_*$ is even in  $y$. 
%Thus, $a_{j,0} = 0$ for $j \leq 3$ and $a_{40}=0$. In addition, reciprocity leads to $a_{jm} = 0$ for all odd $j$ and any $m \geq 0$.
%Reciprocity causes that both the number of extreme points and the Puiseux exponents $\alpha$ differ from the case of propagating super-BICs with order $n=2$. 
For $\delta > 0$, the number and locations of the zeros and extreme points near a SSW for different perturbation types are summarized in Table~\ref{tab:SSW}. Due to reciprocity, $\beta=0$ remains an extreme point of $\mathrm{Im}(k)$ for all perturbation types.

For lossless-SP and lossless-SB perturbations, the results are similar to those of propagating super-BICs with $n=2$. 
For dissipative and gain perturbations,  the number of extreme points depends on the sign of $a_{21}$.
If $a_{21}>0$, there are three extreme points. Otherwise, only a single extreme point remains.
In either case, we have
$
\mathrm{Im}\!\left(k^{\sf Ext}\right) \sim a_{01}\,\delta .
$
If the perturbation is dissipative, then $a_{01}<0$ and all extreme points lie below the real axis. Thus, no zero exists. If the perturbation is gain, then $a_{01}>0$ and all extreme points lie above the real axis. Thus, there are two zeros at two symmetric values of $\beta$, and they correspond to two LTMs.

For $\mathcal{PT}$-symmetric perturbations, if $a_{22} < 0$, there is a maximum at $\beta=0$ with $\mathrm{Im}(k^{\sf Max}) \sim a_{04}\,\delta^4 < 0$. Thus, there is no zero. If $a_{22} > 0$,  there are three extreme points:
two symmetric local maximum points and one local minimum point fixed at $\beta=0$.
The minimum satisfies
$\mathrm{Im}(k^{\sf Min}) \sim a_{04}\,\delta^4 < 0$,
whereas the two maxima satisfy $ \mathrm{Im}(k^{\sf Max}) \sim \xi \delta^4$, where $\xi$ is defined in Appendix A. If $\xi > 0$, we have $\mathrm{Im}(k^{\sf Max})  > 0$  for small $\delta$.
As a result, $\mathrm{Im}(k)$ intersects the real axis at four distinct values of $\beta$  for small positive $\delta$.
Among these four zeros, two correspond to BICs and the other two correspond to LTMs, forming two BIC-LTM pairs.
This behavior is qualitatively different from that near propagating super-BICs and generic BICs, where only two zeros appear. If $\xi < 0$, there is no zero.

\begin{figure}
    \centering
    \includegraphics[width=0.45\linewidth]{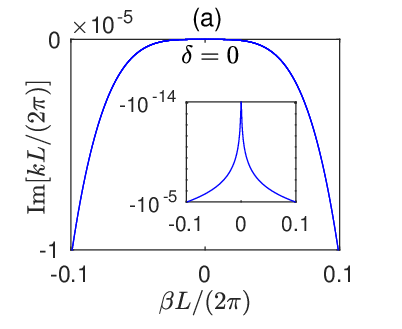}
    \includegraphics[width=0.45\linewidth]{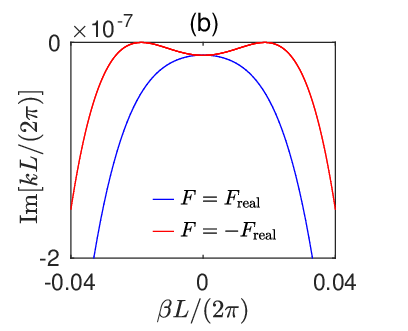} \\
    \includegraphics[width=0.45\linewidth]{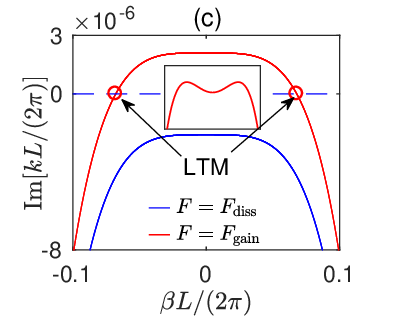} 
     \includegraphics[width=0.45\linewidth]{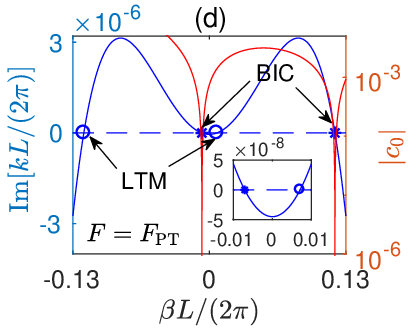}
     
    \caption{
Evolution of the dispersion relation near the SSW. 
(a): $\mathrm{Im}[k(\beta, 0)]$. The inset shows the logarithmic scale. 
(b): $\mathrm{Im}(k)$ for the lossless-SP perturbation with $\delta=0.1$. 
(c): $\mathrm{Im}(k)$ for dissipative (blue) and gain (red) perturbations at $\delta=0.0001$.  The inset shows a zoomed-in view for $\beta \in [-0.008, 0.008] (2 \pi /L)$. 
(d):  $\mathrm{Im}(k)$ for the $\mathcal{PT}$-symmetric perturbation with $\delta=0.1$. Left axis: $\mathrm{Im}(k)$; right axis: radiation coefficient $c_0$ as a function of $\beta$. The inset shows a zoomed-in view for $\beta $ near $0$. 
}
    \label{fig:SSW}
\end{figure}

To validate the theoretical results summarized in Table~\ref{tab:SSW},
we consider the same structure.
For  $W_y=0.25L$, $W_z=0.6771L$,  $W=0.2L$, and the same $\epsilon_1$,
the structure has an SSW  at
$\beta_* = 0$ and $k_* \approx 0.5899\,(2\pi /L)$.
%Fig.~\ref{fig:SSW}(a) shows the imaginary part of the dispersion relation near the SSW for the unperturbed structure.
Fig.~\ref{fig:SSW}(b) shows the case of the lossless-SP perturbation  with $F = \pm F_{\sf real}$ and $\delta=0.1$. When $F = F_{\sf real}$, there is no zero and there is a maximum point, as shown by the blue curve.  For $F = -F_{\sf real}$, two double zeros and three extreme points are observed, as shown by the red curve. 
Fig.~\ref{fig:SSW}(c) illustrates the effect of small dissipative and gain perturbations with $\delta=10^{-4}$.
For the gain perturbation $F = F_{\sf gain}$, three extreme points are observed, as shown in the inset.
The values of \(\mathrm{Im}(k)\) at the three extrema are very close because they share the same leading-order term.
Fig.~\ref{fig:SSW}(d) shows the case of  the $\mathcal{PT}$-symmetric perturbation $F = F_{\sf PT}$ and $\delta=0.1$.
Unlike the propagating super-BICs with $n=2$, in this case, four zeros of $\mathrm{Im}(k)$ are observed.
Among them, two correspond to BICs and the other two correspond to LTMs, as verified by plotting the radiation coefficient as a function of $\beta$.

\subsection{Third-order ASWs}

\begin{table*}[htbp]
    \centering

\caption{Extreme points and zeros of $\mbox{Im}(k)$ for $\delta>0$ near  super ASWs with $n=3$. Constants $\eta_3$, $\chi_1$ and $\chi_2$ are   defined in Appendix~A.}
\label{tab:SuperASW}
    \begin{tabular}{|c|c|c|c|c|c|c|c|}\hline
        \rule{0pt}{3ex}
         \multirow{2}{*}{$F$}   & \multicolumn{4}{c|}{Extreme points} &   \multicolumn{3}{c|}{Zeros}  \\
         \cline{2-8}
         \rule{0pt}{3ex}
           & No. & $\alpha$  & $\beta^{\sf Ext} - \beta_* \sim $ &  $\mbox{Im}(k^{\sf Ext}) \sim $ &  No. & $\alpha$ &  $\beta^{\sf Zero} - \beta_* \sim $ \\[2pt] \hline
\rule{0pt}{6.5ex}
Lossless-SP  & $5$ or $1$ & $\dfrac{1}{2}$ & $ \begin{array}{l} \pm \sqrt{ - \frac{a_{41}}{6a_{60}} } \delta^{\frac{1}{2}},  \\
  \pm \sqrt{ - \frac{a_{41}}{2a_{60}} } \delta^{\frac{1}{2}}, 0 \end{array} $ &  $ \eta_3 \delta^3 $, $0$ &  $3$ or $1$ & $\dfrac{1}{2}$ & $ \beta^{\sf Zero} = \beta^{\sf Max}  $ \\[15pt] \hline
  \rule{0pt}{4ex}
 Lossless-SB  & $5$ or $1$ & $\dfrac{1}{2}$ & $ \pm \sqrt{ \chi_1 \pm \sqrt{\chi_1^2 - \chi_2} } \delta^{\frac{1}{2}}, 0 $ & $ a_{02} \delta^2 $ &  $0$  & $-$ & $ -  $ \\[5pt] \hline 
 \rule{0pt}{4ex}
 Dissipative & $3$ or $1$ & $\dfrac{1}{4}$ & $ \begin{array}{l}  \pm \sqrt[4]{-\frac{a_{21}}{3 a_{60}} } \delta^{\frac{1}{4}}, 
 0  \end{array} $ & $ a_{01} \delta $ & $0$ & $-$  &  $-$  \\[5pt] \hline
 \rule{0pt}{4ex}
 Gain  & $3$ or $1$ & $\dfrac{1}{4}$ & $ \pm \sqrt[4]{-\frac{a_{21}}{3 a_{60}} } \delta^{\frac{1}{4}}, 0$ &  $ a_{01} \delta $ & $2$ & $\dfrac{1}{6}$  & $  \pm \sqrt[6]{-\frac{a_{01}}{a_{60}} }  \delta^{\frac{1}{6}}$ \\[5pt] \hline
 \rule{0pt}{4ex}
 $\mathcal{PT}$  & $3$ or $1$ & $\dfrac{1}{2}$ & $ \begin{array}{l}  \pm \sqrt[4]{ - \frac{a_{22}}{ 3a_{60}} } \delta^{\frac{1}{2}}, 0 \end{array} $ & $ a_{02} \delta^2 $ &  $2$ & $\dfrac{1}{3}$ & $  \pm \sqrt[6]{  - \frac{a_{02}}{a_{60}} } \delta^{\frac{1}{3}}  $ \\[5pt] \hline
 \end{tabular} 
\end{table*}

Antisymmetric standing waves with $n=3$ are the simplest super-BICs with $n > 2$.
For an ASW with $n=3$, $\beta_* = 0$ is a sixth-order zero of $\mathrm{Im}[k(\beta,0)]$, and $\phi_*$ is odd in $y$. %Thus,$a_{j,0}=0$ for $ 0 \le j \le 5,$  and $a_{60}<0.$ In addition, $a_{jm} = 0$ for all odd $j$ and any $m \geq 0$.
The sixth-order zero enables rich splitting scenarios, including the generation of multiple generic BICs and smaller values of $\alpha$.
The number and locations of the zeros and extreme points of $\mathrm{Im}(k)$ for $\delta > 0$ under different types of perturbations are summarized in Table~\ref{tab:SuperASW}.

For lossless-SP perturbations, the results depend on the sign of  $a_{41}$.
If $a_{41}>0$, $\mathrm{Im}(k)$ for $\delta > 0$ has five extreme points.
Among them, three are local maximum points. One local maximum is located at $\beta^{\sf Max} = 0$, and the other two satisfy 
$
\beta^{\sf Max}-\beta_* \sim \pm \sqrt{-{a_{41}}/({2a_{60}})}\,\delta^{1/2}.
$
All three local maximum points coincide with zeros, and they correspond to BICs. This indicates that the super ASW splits into three generic  BICs.
The remaining two extreme points are local minimum points, and satisfy 
$
\beta^{\sf Min}-\beta_* \sim \pm \sqrt{-{a_{41}}/({6a_{60}})}\,\delta^{1/2}
$ and  $
\mathrm{Im}(k^{\sf Min}) \sim  \eta_3 \delta^3 < 0.
$
If $a_{41}<0$, only a single local maximum point remains at $\beta=0$, and  it corresponds to a generic ASW. Note that $a_{41}$ changes sign when $F$ is replaced by $-F$.

For lossless-SB perturbations, $\mathrm{Im}(k)$ for $\delta > 0$ again exhibits either five extreme points or one.
However, all extreme points  lie strictly below the real axis, since $ \mathrm{Im}(k^{\sf Ext}) \sim a_{02} \delta^2 < 0 $. Thus, no zeros are generated.
For dissipative and gain perturbations, the qualitative behaviors are similar to those of SSWs.
However,  Puiseux exponents $\alpha$ in the leading-order expansions of the extreme points and zeros are different.
In addition, if the perturbation profile $F$ is even in $y$, the local extreme point at $\beta=0$ corresponds to a cBIC.

For $\mathcal{PT}$-symmetric perturbations, $\mathrm{Im}(k)$ for $\delta > 0$ has either one or three extreme points, depending on the sign of
$
a_{22}.
$
If $a_{22}>0$, there are two local maximum points and one local minimum point located at $\beta^{\sf Min}=0$.
If $a_{22}<0$, only a single local maximum point remains at $\beta^{\sf Max}=0$.
In both cases, the values of \(\operatorname{Im}(k)\) at these extreme points have the same leading-order scaling,
$
\mathrm{Im}\!\left(k^{\sf Ext}\right) \sim a_{02}\,\delta^2 > 0 .
$
Consequently, two zeros are generated, one corresponding to a BIC and the other to an LTM.
As shown in Appendix~A, the sign of $a_{22}$ is determined by $k_{20} = \partial^2_{\beta}k(\beta_*,0)$. Thus, the number of extreme points of $\mathrm{Im}(k)$ for $\delta > 0$ is determined by the local curvature of the dispersion relation of the unperturbed structure.

\begin{figure}
    \centering
    \includegraphics[width=0.45\linewidth]{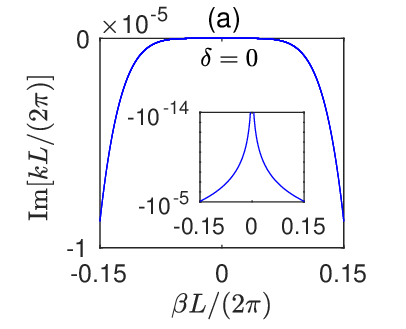}
    \includegraphics[width=0.45\linewidth]{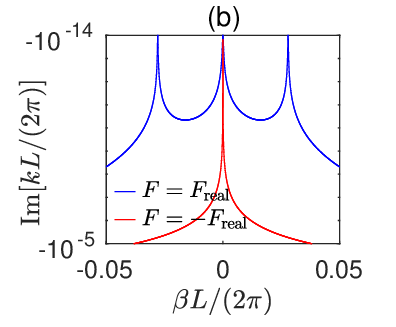} \\
    \includegraphics[width=0.45\linewidth]{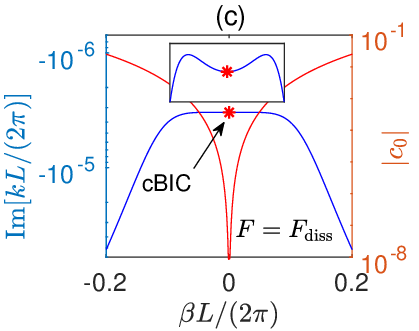} 
     \includegraphics[width=0.45\linewidth]{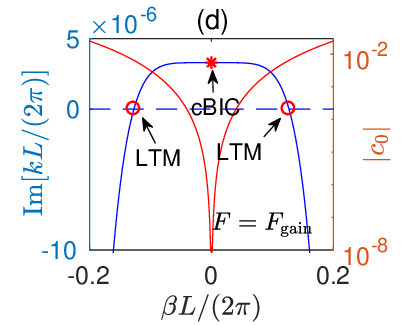}  \\
     \includegraphics[width=0.45\linewidth]{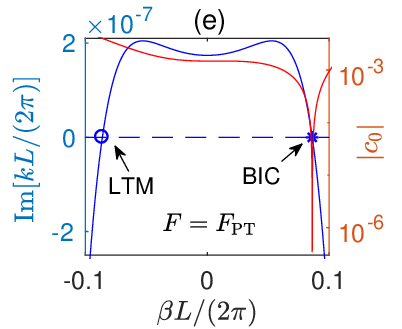}
     
    \caption{
Evolution of the dispersion relation near the super ASW with $n=3$. 
(a): $\mathrm{Im}[k(\beta, 0)]$. The inset shows the logarithmic scale. 
(b): $\mathrm{Im}(k)$ for lossless-SP perturbations at $\delta=1$. 
(c) and (d): $\mathrm{Im}(k)$ for dissipative and gain perturbations at $\delta=0.0001$, respectively.  The inset in (c) shows a zoomed-in view for $\beta \in [-0.05, 0.05] (2 \pi /L)$. 
(e):  $\mathrm{Im}(k)$ for $\mathcal{PT}$-symmetric perturbation with $\delta=0.1$. Left axes of panels (c) - (e): $\mathrm{Im}(k)$; right axes  of panels (c) - (e): radiation coefficient $c_0$ as a function of $\beta$. 
}
    \label{fig:SuperASW}
\end{figure}

To validate the results listed in Table~\ref{tab:SuperASW},
we  consider the same periodic structure.
For $W_z=0.4959L$ and remaining parameters identical to the SSW case,
we find a super ASW of order $n=3$  with $  
k_* \approx 0.7952\,(2\pi /L).
$
Fig.~\ref{fig:SuperASW}(b) depicts the results for the lossless-SP perturbation with $\delta=1$.
For $F=F_{\sf real}$, three double zeros and five extreme points emerge,  as shown by the blue curve.
If the sign of the perturbation is reversed, i.e., $F=-F_{\sf real}$, a single double zero coincides with a local maximum point at $\beta=0$, as indicated by the red curve.
Fig.~\ref{fig:SuperASW}(c) illustrates the effect of the small dissipative perturbation $F = F_{\sf diss}$ and $\delta=10^{-4}$.
Three extreme points are observed, as shown in the inset.
Since the three extrema share the same leading-order term, their values of \(\mathrm{Im}(k)\) are very close.
The local minimum point at $\beta^{\sf Min}=0$ corresponds to a cBIC.
Fig.~\ref{fig:SuperASW}(d) shows the result for the gain perturbation $F = F_{\sf gain}$ and $\delta=10^{-4}$.
In that case, a single local maximum point is observed at $\beta=0$,
which again corresponds to a cBIC.
Fig.~\ref{fig:SuperASW}(e) shows the case of the $\mathcal{PT}$-symmetric perturbation $F = F_{\sf PT}$ and $\delta=0.1$.
Two zeros of $\mathrm{Im}(k)$ are present. We can see that one of them corresponds to a BIC and the other corresponds to an LTM.

\section{Conclusion}
\label{sec:conclusion}
We developed a  framework to analyze the evolution of BICs under small structural perturbations, focusing on the imaginary part of the dispersion relation [i.e., $\mathrm{Im}(k)$]. Using local Taylor expansions and Puiseux series, we determined the number, locations, and leading-order scaling (in powers of $\delta^\alpha$) of zeros and extreme points of $\mathrm{Im}(k)$ for $ \delta > 0$. The framework can be applied to any BIC  and any perturbation type, but we present detailed results only for five different types of BICs and five different types of perturbations.
This approach provides a compact and comprehensive description of BIC-related phenomena, including robustness, splitting, annihilation, and appearance of LTMs.

The local behavior of \(\mathrm{Im}(k)\) near a BIC is governed by the order \(n\), the perturbation type, and the symmetry constraints imposed by reciprocity.
The order sets the maximum number of nearby extreme points and real zeros that may emerge, thereby determining the complexity of possible BIC evolutions.
Reciprocity further reorganizes this evolution by fixing the location of certain extreme points while modifying the multiplicity and Puiseux scaling of the resulting extreme points and zeros.

Although we have only treated five canonical perturbation types, more complicated perturbations can be decomposed into a sequence of these elementary cases. The evolution of the dispersion relation under such a composite perturbation can be analyzed by successive application of the present theory, thereby extending the framework well beyond the five cases considered explicitly.

\section*{Acknowledgments}
The authors acknowledge support from  the National Natural Science Foundation of China (Grant No. 12571384) and the Research Grants Council of Hong Kong Special Administrative Region, China (Project CityU No. 11317622).

\section*{Appendix}
\renewcommand{\theequation}{A\arabic{equation}} % 重新定\UTF{4E49}公式\UTF{7F16}号格式
\setcounter{equation}{0} % 重置\UTF{8BA1}数器

\subsection*{Appendix A: Key coefficients for the Puiseux analysis}

In this appendix, we list the key coefficients \(a_{jm}\) required for the Puiseux analysis and illustrate how \(\alpha\) and \(\beta_1^{\sf Zero}\) are determined for a propagating BIC with \(n=1\).
We consider a BIC that is even in \(z\); the odd case is analogous. We define
\begin{align}
\label{eq:phi_jm}
\phi_{jm}(\mathbf{r})
=\frac{\partial^{j+m}\phi}{\partial\beta^{j}\partial\delta^{m}}
(\mathbf{r};\beta_*,0),  \quad j,m \geq 0.
\end{align}
If, for a given pair $(j,m)$,  $\phi_{jm} \sim q_{jm} e^{ \pm {\sf i} \gamma_* (z \mp d)} $ as $z \to \pm \infty$, we call $q_{jm}$ the radiation coefficient of $\phi_{jm}$.  Here, $\gamma_*  = \sqrt{k_*^2 - \beta_*^2}$.

Let $v_e = \varphi_e e^{{ \sf i} \beta_* y} $ be a diffraction solution corresponding to the BIC as defined in \cite{yuan17}. It satisfies 
$$\varphi_e \sim \overline{\tau} e^{\mp {\sf i} \gamma_* (z \mp d)} + \tau e^{\pm {\sf i} \gamma_* (z \mp d)}, \quad z \to \pm \infty,$$ 
where $|\tau| = 1$. 

\medskip
\noindent
\textbf{I. Propagating BICs with $n=1$ (Table~\ref{tab:PBIC}).}

We have $a_{00}=a_{10}=0$ and $a_{20}=-\sigma_*|q_{10}|^2$, where $\sigma_*=\gamma_* L/k_* > 0$ and $ q_{10} \neq 0$.

For lossless-SP perturbations, $a_{01}=0$,
$a_{11}=-2\sigma_*\overline{q_{10}} {q}_{01}$,  and $a_{02} = -\sigma_* |q_{01}|^2$. We can show that 
\begin{equation} 
    \label{eq:q01}
q_{01} = - \frac{1}{4 {\sf i} \sigma_* \overline{\tau} } \int_{\Omega} \overline{\varphi_e} (2  k_{01} \epsilon_* + k_* F) \phi_* d \mathbf{r},
\end{equation}
where $k_{01} = -k_* \int_{\Omega} F |\phi_*|^2 d \mathbf{r} /2$.
For a generic perturbation profile $F$, $q_{01}\neq 0$. Following the perturbation method in Section~\ref{sec:theory} and  balancing the lowest-order terms, we have $\alpha = 1$ and 
$$
a_{20} \left( \beta^{\sf Zero}_1 + \rho_1 \right)^2  = 0,
$$
where $\rho_1=a_{11}/(2a_{20}) = q_{01}/q_{10}$ is a real number. Thus,  $\beta^{\sf Zero}_1 =  - \rho_1$ is a double root of the leading-order equation. 

For lossless-SB perturbations, $\rho_1$ is generally nonreal, and $a_{11}=2 a_{20} \mbox{Re}(\rho_1)$. The remaining key coefficients are the same as those for lossless-SP perturbations. We have  $\alpha = 1$ and 
$$
a_{20} \left[ \left(\beta^{\sf Zero}_1 \right)^2 + 2 \mbox{Re}(\rho_1) \beta^{\sf Zero}_1 + |\rho_1|^2 \right]  = 0.
$$
If $\mathrm{Im}(\rho_1) \neq 0$, the above equation has no real root.  
The quantity $\eta_1$ in Table~\ref{tab:PBIC} is defined as $\eta_1 = (4 a_{20} a_{02} - a_{11}^2)/{4 a_{20}}$. If $\mathrm{Im}(\rho_1) \neq 0$, then $\eta_1 < 0$.

For dissipative and gain perturbations, we have
\begin{equation}
\label{eq:a01}
a_{01}
=-\frac{k_*}{2}\,
\mathrm{Im}\!\left(
\int_{\Omega}F |\phi_*|^2\,d\mathbf{r}
\right),
\end{equation}
and $a_{11} \neq 0$ for a generic perturbation profile $F$.
Thus, $\alpha=1/2$ and
\[
a_{20} \left(\beta^{\sf Zero}_1 \right)^2+a_{01}=0.
\]
For dissipative perturbations, $\mathrm{Im}(F) \geq 0$, so $a_{01}\leq0$. Assuming a nontrivial overlap between the dissipative region and the BIC field, we have $a_{01} < 0$. Thus, the above equation has no real
root.
For gain perturbations, under the corresponding nontrivial-overlap assumption, we have $a_{01}>0$, and there are two
real roots
$\beta^{\sf Zero}_1=\pm\sqrt{-a_{01}/a_{20}}$.

For $\mathcal{PT}$-symmetric perturbations, $F$ is purely imaginary and
odd in $y$. In this case, we have $a_{01}=a_{11}=0$, $a_{12} \neq 0$, and $a_{02}=\sigma_*|q_{01}|^2$. Since $F$ is odd in $y$, we have  $k_{01}  = 0$ and 
$$ q_{01} = - \frac{k_*}{4 {\sf i} \sigma_* \overline{\tau} } \int_{\Omega} \overline{\varphi_e}  F \phi_* d \mathbf{r}. $$
For generic $F$, $q_{01} \neq 0$ and $a_{02} > 0$.
In this case, we have  $\alpha=1$ and 
\[
a_{20} \left(\beta^{\sf Zero}_1 \right)^2+a_{02}=0,
\]
which again yields two real roots.

The extrema can be determined analogously from
$ \partial_\beta\operatorname{Im}(k)=0$.

\medskip
\noindent
\textbf{II. Super-BICs.}  

In the following, we list only the key nonzero coefficients \(a_{jm}\) required for the Puiseux analysis. For each perturbation type and each fixed \(m\), the coefficients \(a_{jm}\) are ordered by increasing \(j\); up to the largest listed value of \(j\), every unlisted coefficient is understood to vanish. For example, for propagating super-BICs with \(n=2\) under lossless-SP perturbations, \(a_{01}=a_{11}=0\), so \(a_{21}\) is the first nonzero coefficient in the sequence \(\{a_{j1}\}_{j\geq0}\). For each perturbation class, the listed coefficients are nonzero for generic choices of the perturbation profile \(F\).

\smallskip
\noindent \textbf{Propagating super-BICs with $n=2$ (Table~\ref{tab:SuperPBIC}):} 

For all perturbation types, $ a_{40} = -{\sigma_*}|q_{20}|^2 / 4 $ and $q_{20} \neq 0$.

\begin{itemize}[leftmargin=2cm]
    \item[Lossless-SP:] $a_{21} = -\sigma_*\,\overline{q_{20}} q_{01}$,  $a_{02} = -\sigma_* |q_{01}|^2$, $a_{12} \neq 0$.
    \item[Lossless-SB:] $a_{21} = -\sigma_*\,\mbox{Re}(\overline{q_{20}} q_{01})$,  $a_{02} = -\sigma_* |q_{01}|^2$, $a_{12} \neq 0$.
\item[Dissipative:]  $a_{01} < 0$, $a_{11} \neq 0$.
\item[Gain:] $a_{01} > 0$, $a_{11} \neq 0$.
\item[$\mathcal{PT}$:]  $a_{02} = \sigma_* |q_{01}|^2$, $a_{12} \neq 0$.
\end{itemize}  
In the lossless-SP case, $\overline{q_{20}} q_{01}$ is real.
In the lossless-SB case, $\overline{q_{20}} q_{01}$ is generally nonreal, and $\eta_2 = ({4 a_{40} a_{02} - a^2_{21}})/({4 a_{40}}) < 0$ provided that $\mathrm{Im}(\overline{q_{20}} q_{01}) \neq 0$.

\smallskip
\noindent \textbf{SSWs with $n=2$ (Table~\ref{tab:SSW}): } 

For all perturbation types, $ a_{40} = -{\sigma_*}|q_{20}|^2 / 4 $ and $q_{20} \neq 0$.

\begin{itemize}[leftmargin=2cm]
    \item[Lossless-SP:] $a_{21} = -\sigma_*\,\overline{q_{20}} q_{01}$,
 $a_{02} = -\sigma_* |q_{01}|^2$.
 \item[Lossless-SB:] $a_{21} = -\sigma_*\,\mbox{Re}( \overline{q_{20}} q_{01})$,
 $a_{02} = -\sigma_* |q_{01}|^2$.
\item[Dissipative:] $a_{01} < 0$,  $a_{21} \neq 0$.
\item[Gain:] $a_{01} > 0$,  $a_{21} \neq 0$.
\item[$\mathcal{PT}$:]  
$a_{22} = \sigma_* \left(|q_{11}|^2 -\overline{q_{20}} q_{02}/2 \right)$,
$a_{04} = -{\sigma_*} |q_{02}|^2/4 $.
\end{itemize}
In the lossless-SP case, $\overline{q_{20}} q_{01}$ is real.
In the lossless-SB case, $\overline{q_{20}} q_{01}$ is generally nonreal,
and $\eta_2 = ({4 a_{40} a_{02} - a^2_{21}})/({4 a_{40}}) < 0$ provided that $\mathrm{Im}(\overline{q_{20}} q_{01}) \neq 0$. In the $\mathcal{PT}$-symmetric case, $\overline{q_{20}} q_{02}$ is real. We define
$ \xi = (4 a_{40} a_{04} - a_{22}^2)/(4 a_{40}) $,
$\xi_1 = - a_{22}/(2a_{40})$, and $ \xi_2 = a_{04}/a_{40}.$

\smallskip
\noindent \textbf{ASWs with $n=3$ (Table~\ref{tab:SuperASW}):}

For all perturbation types, $ a_{60} = -{\sigma_*}|q_{30}|^2 / 36$ and $q_{30} \neq 0$.

\begin{itemize}[leftmargin=2cm]
    \item[Lossless-SP:] 
$a_{41} = -\sigma_*\,\overline{q_{30}} q_{11} / 3$, 
$a_{22} = -\sigma_* |q_{11}|^2$.
\item[Lossless-SB:] 
$a_{41} = -\sigma_*\,\mbox{Re}(\overline{q_{30}} q_{11}) / 3$, $a_{02} = -\sigma_* \ |q_{01}|^2, $
$a_{22} = -\sigma_* |q_{11}|^2$.
\item[Dissipative:]
$a_{01} < 0$,  
$a_{21} \neq 0$.
\item[Gain:]
$a_{01} > 0$, 
$a_{21} \neq 0$.
\item[$\mathcal{PT}$:]
$a_{02} = \sigma_* |q_{01}|^2$,  
$a_{22} = 2 \sigma_* {k_{20}} |q_{01}|^2 / k_*$.
\end{itemize}
In the lossless-SP case, $\overline{q_{30}} q_{11}$ is real, and $\eta_3 = - a_{41}^3/({54 a^2_{60}}) $.
In the lossless-SB case, $\overline{q_{30}} q_{11}$ is generally nonreal. We define $\chi_1 = - a_{41}/(3a_{60}) =  -4  \mbox{Re}(q_{11}/q_{30})$ and $\chi_2 = a_{22}/(3a_{60}) = 12  |q_{11}/q_{30}|^2$.

\subsection*{Appendix B: Distinguishing BICs from LTMs}
\renewcommand{\theequation}{B\arabic{equation}}
\setcounter{equation}{0}

Under $\mathcal{PT}$-symmetric perturbations, a real zero of $\mathrm{Im}(k)$ may correspond to  a BIC or an LTM.  
In this appendix, we present a procedure to distinguish these two possibilities by examining the asymptotic behavior of the associated mode field.  
As an example, we focus on a propagating super-BIC with $n=2$, where $\mathrm{Im}(k)$ has two zeros for $\delta > 0$.

For $\delta > 0$, we expand the wave field corresponding to a zero,  $\phi^{\sf Zero} = \phi[\mathbf{r}; \beta^{\sf Zero}(\delta), \delta] $, as a  Puiseux series:
\begin{eqnarray}
    \label{eq:exp_phi_zero}
\phi^{\sf Zero} 
= \phi_* + \phi_1^{\sf Zero}\,\delta^{\alpha}
+ \phi_2^{\sf Zero}\,\delta^{2\alpha}
+ \ldots .
\end{eqnarray}
If  every coefficient $\phi_p^{\sf Zero}$, $p \ge 1$,  decays to zero as $z \to \pm \infty$, then $\phi^{\sf Zero}$  corresponds to a BIC.  
Otherwise, if at least one $\phi_p^{\sf Zero}$ exhibits non-decaying behavior at infinity, $\phi^{\sf Zero}$ corresponds to an LTM.

Substituting the Puiseux expansion of \(\beta^{\sf Zero}\) in Eq.~\eqref{eq:exp_beta_zero} into the Taylor expansion of \(\phi(\mathbf r;\beta,\delta)\) about \((\beta_*,0)\), and collecting equal powers of \(\delta\), we can determine the coefficients \(\phi_p^{\sf Zero}\) recursively.
Since \(\alpha=1/2\) for the present case, the first two coefficients are
$
\phi_1^{\sf Zero} = \beta_1^{\sf Zero}\,\phi_{10},
$
and
\[
\phi_2^{\sf Zero}
= \beta_2^{\sf Zero}\,\phi_{10}
+ \frac{1}{2} \left( \beta_1^{\sf Zero} \right)^2 \phi_{20}
+ \phi_{01}.
\]
For a propagating super-BIC with $n=2$, $\phi_{10} \to 0$ as $z \to \pm \infty$.  
Consequently, $\phi_1^{\sf Zero} \to 0$, while
\[
\phi_2^{\sf Zero}
\sim
\left[
\frac{1}{2} \left( \beta_1^{\sf Zero} \right)^2 q_{20}
+ q_{01}
\right]
\mathrm{e}^{\pm {\sf i} \gamma_* (z \mp d)},
\quad z \to \pm \infty .
\]

According to Table~\ref{tab:SuperPBIC}, the leading coefficients of the two zeros are
$
\beta_1^{\sf Zero}
= \pm \left( -\frac{a_{02}}{a_{40}} \right)^{\frac{1}{4}}
= \pm \sqrt{2|\chi|},
$
where
$
\chi = {q_{01}}/{q_{20}} 
$. The relevant symmetry relation implies that $\chi$ is real. For generic $F$, $q_{01} \neq 0$ and $\chi \neq 0$.  
Therefore, $$
\phi_2^{\sf Zero}
\sim
 q_{20}(|\chi|+\chi) e^{\pm {\sf i} \gamma_* (z \mp d)}, \quad z \to \pm \infty.$$
If $\chi>0$, then $|\chi|+\chi = 2 \chi \neq 0$. Hence, \(\phi_2^{\sf Zero}\) contains a nonzero outgoing component, and both zeros  correspond to LTMs.
If $\chi<0$, then $|\chi|+\chi = 0$ and $\phi_2^{\sf Zero}$ decays to zero as $z \to \pm \infty$. 
Using the recursive equations, one can prove by induction that every higher-order coefficient \(\phi_p^{\sf Zero}\), \(p\geq3\), also decays as \(z\to\pm\infty\). Therefore, both zeros correspond to BICs.

\end{document}